 \documentstyle[prl,aps,multicol,epsfig]{revtex}

\begin{document}
\draft
\widetext

 \title{Theory for the Interdependence of High-T$_c$ Superconductivity\\
and Dynamical Spin Fluctuations}

\author{ S. Grabowski, J. Schmalian, M. Langer, and K. H. Bennemann}
\address{ Institut f\"ur Theoretische Physik,
 Freie Universit\"at Berlin, Arnimallee 14, 14195 Berlin , Germany}

\date{June 29, 1995}
\maketitle

\widetext
\begin{abstract}
\leftskip 54.8pt
\rightskip 54.8pt
The doping dependence of the superconducting state for the 2D
one-band Hubbard Hamiltonian is determined. By using an
Eliashberg-type theory, we find that the gap function $\Delta_{\bf k}$
has a $d_{x^2-y^2}$ symmetry in momentum space and
T$_c$ becomes maximal for $13 \; \%$ doping. Since we determine
the dynamical excitations directly from real frequency axis calculations,
we obtain new structures in the angular resolved density of states related
to the occurrence of {\it shadow states} below T$_c$. Explaining the anomalous
behavior of photoemission and tunneling  experiments in the cuprates, we
find a strong interplay between $d$-wave superconductivity
and dynamical spin fluctuations.

\end{abstract}


\begin{multicols}{2}

\narrowtext
Despite important progress, the nature of the superconducting pairing
mechanismof the High-T$_c$-materials is still controversial. Due to
their unconventional behavior in the normal as well as in the
superconducting state various solely electronic pairing mechanism were
proposed~\cite{dagotto}. During the last years the symmetry of the
superconducting order parameter in momentum space was studied
intensively, because it probably holds the key for an understanding of the
High-$T_c$ systems and the role of spin fluctuations within the
cuprates~\cite{schriefer}. Recent angular resolved photoemission
(ARPES) experiments and phase sensitive measurements of the gap
function~\cite{shen,ding,kirt2,wollman,math} clearly favor
a $d_{x^2-y^2}$ or an anisotropic $s_{x^2+y^2}$ symmetry of the gap
in momentum space over an isotropic $s$-wave scenario. Moreover the
maximum of $T_c$ upon doping,  the occurrence of an additional dip in
the ARPES spectra of Bi$_{2}$Sr$_{2}$CaCu$_{2}$O$_{8+\delta}$
found by Dessau {\it et al.}~\cite{dessau2} and the dip at
$\omega = \pm 3 \Delta$ in superconductor-insulator-superconductor (SIS)
tunneling measurements~\cite{mandrus} are important experiments that
may be significant clues for the pairing interaction. In particular, the
recent observation of shadows of the Fermi surface (FS) in the paramagnetic
state by  Aebi {\it et al.}~\cite{aebi} and their interpretation in terms of
short-range antiferromagnetic correlations could be the {\it Smoking Gun}
of a spin  fluctuation pairing mechanism. Therefore, the behavior of the
shadow states below $T_c$ is of great importance for an understanding
of the superconductivity in the High-$T_c$ systems.

Theoretically, a favorite model to study a purely electronically
mediated pairing interaction in the CuO$_2$ plane is the 2D one-band
Hubbard Hamiltonian. Recently, it was demonstrated within an
Eliashberg-type theory based on the  spin-fluctuation mechanism,
that there exists a superconducting ground state below
$T_c \approx 0.02 t$ with a $d$-wave symmetry of the order
parameter~\cite{mont,pao,pines}. Despite these interesting results,
the dynamical properties and their relation to the strong antiferromagnetic
correlations are far from  being  understood, because the relevant strong
coupling equations were solved on the imaginary
frequency axis which gives no direct access to the dynamical
excitation spectrum. A first  step to determine the excitation spectrum
from a real axis calculation was achieved in an important study by
 Dahm {\it et al.}~\cite{dahm}.

In this Letter, we present results for the excitation spectrum in the
superconducting state and explain the anomalous behavior of the
ARPES spectra and tunneling measurements. In addition, we
determine the doping dependence of the superconducting state
and investigate the interdependence of the superconducting phase
and the strong antiferromagnetic correlations within the cuprates.
Our theory is based on a strong coupling Eliashberg-type approach
for the one-band Hubbard Hamiltonian with nearest neighbor
hopping integral $t=0.25 \, {\rm eV}$, bare  dispersion
$\varepsilon_0 ({\bf k})=-2t[\cos(k_x)+\cos(k_y)]-\mu$ with chemical
potential $\mu$, and local Coulomb repulsion $U=4t$. Since we are
using our new numerical method~\cite{shadow} for the self
consistent summation of all bubble and  ladder diagrams
(fluctuation exchange approximation, FLEX~\cite{flex1}) on the
real frequency axis, we obtain directly the quasi particle excitation
spectrum below T$_c$.  The superconducting state is treated in the
Nambu formalism where the diagonal and off-diagonal Greens function
in the matrix notation can be expanded in terms of the
Pauli matrices $\hat \tau_{i}$ ($i = 0,1,3$):
\[
\hat G({\bf k},\omega)   =
\frac{\omega Z({\bf k},\omega) \hat \tau_{0} - (\varepsilon_0 ({\bf k})
+ \chi({\bf k},\omega)) \hat \tau_{3} + \phi({\bf k},\omega)\hat \tau_{1} }
{(\omega Z({\bf k},\omega))^2 - (\varepsilon_0 ({\bf k})
+ \chi({\bf k},\omega))^2 - (\phi({\bf k},\omega))^2}\,.
\]
Here, $\omega (1-Z({\bf k},\omega))$ and  $\chi({\bf k},\omega)$ are
the diagonal expansion   coefficients of the electronic self energy matrix,
 whereas  $\phi({\bf k},\omega) = \Delta ({\bf k},\omega) Z({\bf k},\omega)$
is the anomalous self energy and $\Delta ({\bf k},\omega)$ the gap
function~\cite{details}. Hence, we calculated these functions by solving
the Eliashberg equations within the FLEX approximation~\cite{imag},
were the ${\bf k}$ and $\omega$ dependent pairing interaction and the
Greens function $\hat G({\bf k},\omega)$ are calculated self
consistently~\cite{para}.

In Fig. 1 we show the density of states $\rho(\omega)$ in the superconducting
 phase which is always of d$_{x^2-y^2}$ pairing symmetry for different
doping concentrations $x=1-n$ and for $T = 63 \; K$. Here, $n$ is the
occupation number per site. For intermediate doping ($x = 0.14$)
a pronounced superconducting gap appears. Due to the d-wave pairing
symmetry, states in the gap are clearly visible. For larger doping
($x = 0.18$) the antiferromagnetic correlations are
smaller and consequently the pairing interaction and the
superconductivity becomes weaker,
while for smaller doping ($x = 0.12$) the magnetic correlations are
more dominant and we find
a sharp increase of the quasi particle scattering. This leads to a
 loss of metalicity and a
weaker superconducting gap. In addition, a pseudogap as a precursor
of the Mott-Hubbard splitting
becomes visible in $\rho(\omega)$.

\begin{figure}
\vskip -1.5cm
\centerline{\epsfig{file=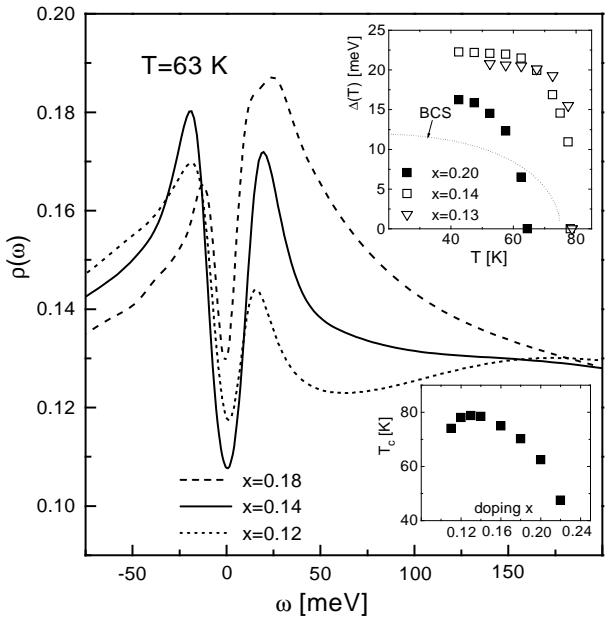,width=9cm,height=12cm}}
\vskip -2.0cm
\caption{ Density of states $\rho (\omega)$ for three doping concentrations.
In the upper inset $\Delta (T)$ is plotted for various doping values.
The dotted line is the corresponding BCS curve with an arbitrary chosen
$T_c = 75 \; K$. In the lower inset the doping dependence of
$T_c$ obtained from $\Delta(T_c)=0$ is shown.}
\label{fig1}
\end{figure}
In the upper inset of Fig. 1 we present data for the superconducting gap
 function $\Delta (T)$
estimated from the distance between the maxima of $\rho(\omega)$ for
different doping values.
For all doping concentrations  the gap opens much faster below $T_c$
than in the corresponding
BCS case (dotted lines) and saturates quickly. This  is due to the shift
of spectral weight
in the pairing interaction to higher energies induced by the opening of
the superconducting
gap and therefore to strong feedback effects~\cite{mont}. Furthermore,
we find in agreement
with Raman scattering measurements that the gap opens more rapidly
when $x$ is decreased
{}~\cite{raman}. Interestingly, the tendency that the gap function
$\Delta (T)$ saturates at higher
values of $\Delta (0)$ when the doping is decreased is reversed between
$x = 0.14$ and
$x = 0.13$. This behavior reflects the increasing stiffness of the system
 due to the oncoming
antiferromagnetic phase transition and to the reduction of the charge
carrier concentration at
the Fermi level resulting from the formation of a pseudogap. Therefore,
the interplay between
antiferromagnetic correlations and superconductivity leads to an optimal
doping concentration.
This is  demonstrated in the lower inset of Fig. 1, where we present in
qualitative agreement
with the experimental observation the doping dependence of $T_c$
becoming maximal for $x = 0.13$
and $T_c = 79 \; K$.

Recently, we presented corresponding results for the normal
state~\cite{shadow} and
argued that the observation of shadows of the FS shifted by
${\bf Q}=(\pi,\pi)$, by
Aebi {\it et al.}~\cite{aebi} is due to strong magnetic correlations
that can be explained
in our framework without a staggered antiferromagnetic moment
and with a small correlation
length of $2.5$ lattice spacings. Now, we propose that these
shadow states also exist
below $T_c$ and that their intensity is even increased compared
to the paramagnetic state.

\begin{figure}
\vskip -0.5cm
\centerline{\epsfig{file=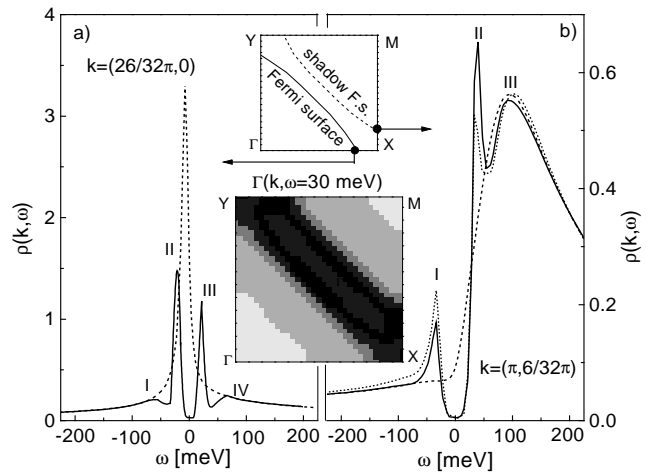,width=9cm,height=12cm}}
\vskip -4.5cm
\caption{ Spectral density $\rho ({\bf k},\omega)$ for two ${\bf k}$
values (see upper inset).
(a) $\rho ({\bf k},\omega)$ for $x=0.16$ and $T = 43 \; K$
(solid line) compared with
$T = 75 \; K$ (dashed line). Note, $T_c ( x=0.16 ) = 75 \; K$.
(b) Solid and dashed lines as in (a) and dotted line for $x=0.14$
and $T = 43 \; K$, where
$T_c ( x=0.14 ) = 78.5 \; K$. The labeled peaks are discussed
in the text. The lower inset shows
the ${\bf k}$ dependence of the scattering rate
$\Gamma ({\bf k},\omega)$ for $\omega = 30$ meV
where the intensities are represented in a linear grey scale.
Note, the largest values of
$\Gamma ({\bf k},\omega)$ are along the main FS and its shadow.}
\label{fig2}
\end{figure}
In Fig. 2 the spectral density $\rho ({\bf k},\omega)$ for
 $x = 0.16$ and
$T = 43 \; K$ is compared with $\rho ({\bf k},\omega)$
in the normal state.
In Fig. 2(a) and 2(b), we plot $\rho ({\bf k},\omega)$ for
${\bf k} = (26/32 \pi,0)$
close to the FS and for ${\bf k} = (\pi,6/32 \pi)$. For the
latter momentum one expects
in the normal phase the shadow states. As was shown in
 Ref.~\cite{shadow} these states
are clearly visible for smaller doping ($x = 0.12$). In distinction
for $x = 0.14$ only a
weak maximum and for $x = 0.16$ no shadow states are
observable for this ${\bf k}$ point above
T$_c$.

\begin{figure}
\vskip -1.5cm
\centerline{ \epsfig{file=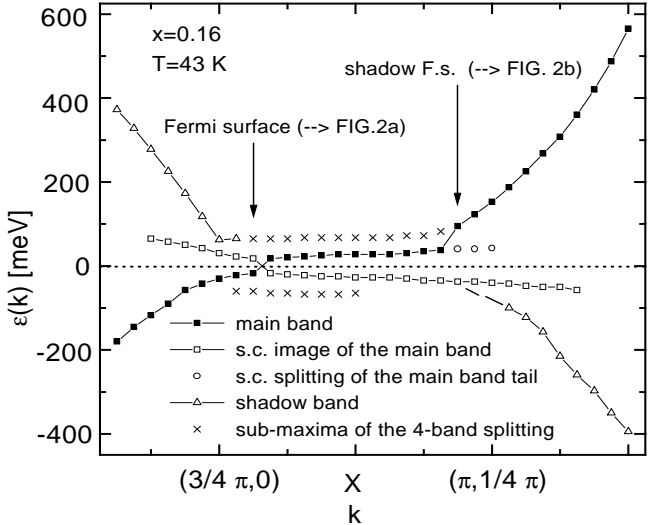,width=9.5cm,height=12cm}}
\vskip -3cm
\caption{Quasi particle dispersion below T$_c$ along the high
symmetry line
$\Gamma \to X \to M $ in the neighborhood of the $X$ point.
The different symbols
represent the dominant contribution to the peaks in
$\rho ({\bf k},\omega)$.}
\label{fig3}
\end{figure}
In Fig. 2(a), the peak which is slightly below the Fermi level
exhibits a superconducting
splitting that is almost symmetrical with respect to $\omega = 0$.
Besides the main peaks two
dips at $\omega \approx \pm 40$ meV are clearly visible. This is
related to a strongly enhanced
quasi particle decay as will be discussed below.

In Fig. 2(b) the main peak in the normal state is well above the
Fermi level. However, a
superconducting splitting can be seen below T$_c$. Pronounced
quasi particle decay processes in
analogy to Fig. 2(a) and the superconducting splitting of the
main band tail for
$| \omega | < | \Delta_k |$ lead to the significant dip structure
 for $\omega > 0$. Most
interestingly, these structures can only be found for ${\bf k}$
states near the FS and the
corresponding shadow states, if $\Delta_k \neq 0$. This results
from an anomalous ${\bf k}$
dependence of the quasi particle scattering rate
$\Gamma ({\bf k},\omega) = {\rm Im} (\omega Z ({\bf k},\omega)) /
 {\rm Re Z} \; ({\bf k},\omega)$
shown in the lower inset of Fig. 2 for $\omega = 30$ meV. Its maxima
occur close to the FS but
also at the position of its shadow with almost equal intensities leading
to a suppression
of the spectral weight. We expect this result to be also of importance
 for the unusual transport
properties of the cuprates. In addition, it is highly interesting that the
dip structure in
Fig. 2(b) is in good agreement with the corresponding ARPES
measurements by Dessau {\it et al.}
in the BISCO system~\cite{dessau2,trans}. By decreasing the
doping to $x = 0.14$ the distance
between the quasi particle peak in the normal state and the Fermi
level increases. Therefore,
the spectral weight of the superconducting quasi particles in peak
$II$ decreases compared to
$x = 0.16$, which one also would expect for the superconducting
 image peak $I$. However, one
clearly sees an increase of peak $I$. In order to show that this
observation is due to the
formation of superconducting quasi particles within the shadow
states, we analyzed the quasi
particle dispersion in the superconducting phase.

In Fig. 3, we present this dispersion $\varepsilon ({\bf k})$
obtained from the maxima of
$\rho ({\bf k},\omega)$. The filled squares indicate the
main-band whereas the open squares
are due to its superconducting splitting. Note that both exhibit
a flat dispersion within a large
${\bf k}$ range. This is due to the flat bands in the paramagnetic
state and
the fact that only the spectral weight below
$ | \omega | \approx | \Delta_{\bf k} |$ is
splitted in the superconducting phase. Furthermore, the
states indicated by open triangles
are the shadow states shifted by ${\bf Q}$ with respect to the
 main band. In the paramagnetic
state and for $x = 0.16$ these shadow bands are also existent,
but their intensities are much
weaker. Therefore, a clear amplification of the shadows below
T$_c$ occurs. Moreover, we indicate with crosses the maxima
caused by the formation of the dip
structure. They result, similar to the shadow states, from strong
quasi particle decay. For that
reason it is tempting to assume that these structures are the
continuation of the shadow
band structure below T$_c$ for a system with purely dynamical
antiferromagnetic correlations.
The four band splitting at the FS is due to the fact that the
frequency dependence of the
quasi particle decay rate of a given band determines that
of its superconducting image state.
Since both peaks behave symmetrically, we find the significant
four band splitting at the FS.
Finally, this amplification of the shadow states explains the
anomalous doping dependence of
the spectral density shown in Fig. 2(b). Although the distance
of the peak $II$ and the Fermi
level increases upon doping, the crossing point of the shadow
states and the Fermi level is
shifted towards ${\bf k} = (\pi,6/32 \pi)$. The anomalous
behavior of peak $I$ results from
an increase of its shadow state contribution for smaller
doping.

{}From the effective interaction of the quasi particles, we find
that the antiferromagnetic
coupling in the superconducting state is even stronger compared
to the situation above $T_c$.
Although the magnetic correlation length ($\approx$ 3 lattice
spacings) increases by less that
one lattice spacings compared to the paramagnetic phase, a
clear increase of the effective
interaction ($ \approx 15 \%$) occurs. Due to the retarded
coupling of singlet Cooper pairs
between nearest neighbor sites, the antiferromagnetic
correlations are stabilized on the
corresponding time scale. These results indicate not only that
the d-wave superconductivity is
caused by antiferromagnetic correlations, but that the presence
of the superconducting state
is also enhancing the short range antiferromagnetic fluctuations.
In order to demonstrate
further consequences of our theory we also calculated the
tunneling conductance for
SIN and SIS junctions, because SIS in-plane tunneling
measurements exhibit an interesting
dip structure at $\omega = \pm 3 \Delta$ whose temperature
dependence is much stronger
compared to the main gap.

\begin{figure}
\vskip -1.4cm
\centerline{\epsfig{file=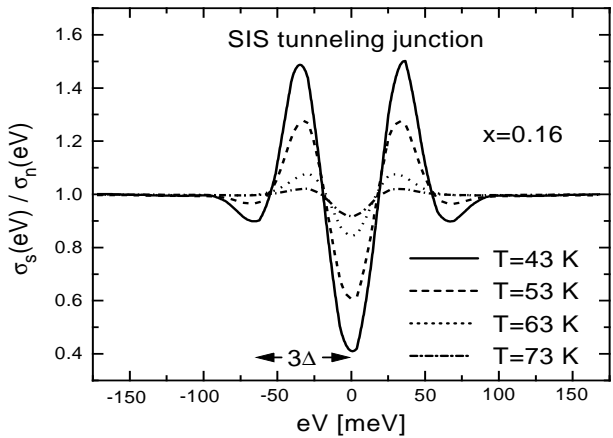,width=9cm,height=11cm}}
\vskip -3.5cm
\caption{SIS tunneling conductance $\sigma_s (eV)$ for
$x=0.16$ normalized by the
paramagnetic $\sigma_n (eV)$ for different temperatures.
Note the strong temperature
dependence of the additional dip structure at
$\pm 3 \Delta$.}
\label{fig4}
\end{figure}
In Fig. 4 we present results for the SIS in-plane
tunneling conductance $\sigma_s (eV) = d I(eV)/d V$~\cite{mandrus}
using
\[
I(eV)\propto  \int_{-\infty}^{\infty} d\omega \rho(\omega)   \rho(\omega+eV)
(f(\omega)-f(\omega+eV))
\]
for the tunneling current. Here, $e$ is the unity charge,
$f(\omega)$ the Fermi function,
and $V$ the applied bias
voltage. By inserting our self-consistently calculated
$\rho(\omega)$, we
find an excellent agreement for both the temperature
dependence and the position
of the dip with the experimental data. Note that in
agreement with SIN measurements,
no pronounced dip in the corresponding tunneling can
be observed (not shown).
It is important to point out that
our interpretation of the $3 \Delta$ dip structure in terms
of a strong increase of the
scattering rate $\Gamma ({\bf k},\omega)$ at the shadow
states is in contrast to the
explanation of Ref.~\cite{coffey}.
Therein its authors argued that the dip in the SIS
measurements is caused by a sudden
increase of $\Gamma ({\bf k},\omega)$ at wave vectors
$k_x= \pm k_y$ where the gap
vanishes.

In conclusion, we calculated the superconducting properties
of the 2D Hubbard model.
We found that the order parameter has a $d_{x^2-y^2}$
symmetry in momentum space and that
$T_c$ is maximal for the doping concentration $x=0.13$.
It was demonstrated for the first time
that the shadow states observed by Aebi~\cite{aebi} are
still present and even enhanced below
$T_c$. We obtain a four-fold splitting of the quasi particle
dispersion at the FS and a dominant dip structure in the
spectral density
$\rho ({\bf k},\omega)$ next to the shadows of the FS.
These new
structures are caused by purely dynamical antiferromagnetic
spin fluctuations.
In view of the recent phenomenological model for a
spin fluctuation induced anisotropic
$s$-wave pairing symmetry in bilayers~\cite{licht}
we believe that in our theory
a similar interplay of antiferromagnetic and
superconducting excitations occurs
for multi-layer systems too. From  our spectral density
we calculate the in-plane SIS tunneling
conductance  and demonstrate that the  dip  in the ARPES
and the $3 \Delta$ dip in the
tunneling characteristics arise for the same physical reason.
Therefore, we believe that these
important experiments and the observation of shadows of
the FS by Aebi {\it et al.} can be
explained within our theory. All this sheds new light on the
superconducting pairing mechanism
and supports the role of spin fluctuations for the high
T$_c$ in the cuprates.
%
%
%
%

\end{multicols}

\begin{references}
%
\bibitem[1]{dagotto}

E. Dagotto, Rev. Mod. Phys. {\bf 66}, 763 (1994) and references therein.

\bibitem[2]{schriefer}

J.R. Schrieffer, Solid State Comm. {\bf 92} 1-2, 129 (1994).

\bibitem[3]{shen}

Z.X. Shen, W.E. Spicer, D.M. King, D.S. Dessau, and B.O. Wells,
Science {\bf 267}, 343 (1995).

\bibitem[4]{ding}
H. Ding {\it et al.}, Phys. Rev. Lett. {\bf 74}, 2784 (1995).

 M. Randeria, T. Takahashi,

\bibitem[5]{kirt2}

J.R. Kirtley {\it et al.}, Nature {\bf 373}, 225 (1995).


\bibitem[6]{wollman}

D.J. Van Harlingen, Rev. Mod. Phys. {\bf 67}, 515 (1995) and
references therein.


\bibitem[7]{math}

A. Mathai {\it et al.}, Phys. Rev. Lett. {\bf 74}, 4523 (1995).


\bibitem[8]{dessau2}

D.S. Dessau {\it et al.}, Phys. Rev.  Lett {\bf  66}, 2160 (1991).


\bibitem[9]{mandrus}

D. Mandrus, L. Forro, D. Koller, and Mihaly, Nature {\bf 351}, 460 (1991).

\bibitem[10]{aebi}

P. Aebi {\it et al.}, Phys. Rev. Lett. {\bf 72}, 2757 (1994).


\bibitem[11]{mont}

P. Monthoux and D.J. Scalapino, Phys. Rev. Lett. {\bf 72}, 1874 (1994).

\bibitem[12]{pao}

C.H. Pao and N.E. Bickers, Phys. Rev. Lett. {\bf 72}, 1870 (1994).

\bibitem[13]{pines}

P. Monthoux and D. Pines, Phys. Rev. Lett. {\bf 69}, 961 (1992).

\bibitem[14]{dahm}

T. Dahm and  L. Tewordt, Phys. Rev.  Lett {\bf  74}, 793 (1995).

\bibitem[15]{shadow}

M. Langer, J. Schmalian, S. Grabowski, and K.H. Bennemann, preprint.

\bibitem[16]{flex1}
N. E. Bickers, D. J. Scalapino, Ann. Phys. (N.Y.) {\bf 193}, 206 (1989).

\bibitem[17]{details}

Details will be published elsewhere.

\bibitem[18]{imag}

Note, a version of the Eliashberg equations on the imaginary axis is given
in Ref.~\onlinecite{mont}.

\bibitem[19]{para}

The calculations are performed on a ($64 \times 64$) square lattice.
We use 4096 equally spaced energy points in the interval $[-30 t,30 t]$,
leading to a low energy resolution of $0.014 t$
(see Ref.~\onlinecite{shadow}).

\bibitem[20]{raman}

T. Staufer, R. Nemetschek, R. Hackl, P. M\"uller, and H. Veith,
Phys. Rev. Lett. {\bf 68}, 1069 (1992).

\bibitem[21]{trans}

Note that for the comparison with the experiments one has to
perform a particle hole
transformation, i.e. an exchange of occupied and unoccupied
states and of momenta
${\bf k}$ and ${\bf k}+{\bf Q}$.

\bibitem[22]{coffey}

D. Coffey and L. Coffey, Phys. Rev.  Lett {\bf  70}, 1529 (1993).

\bibitem[23]{licht}

A.I. Liechtenstein, I.I. Mazin, and O.K. Andersen, Phys. Rev.
Lett {\bf  74}, 2303 (1995).

\end{references}
\end{document}